\documentclass[final,3p]{CSP}

\usepackage{amsmath, amssymb, amsfonts, amsthm, latexsym, bm, mathdots, accents}
\usepackage{graphicx}
\usepackage{float}
\usepackage{tikz}
\usetikzlibrary{matrix,positioning,cd}
\usepackage{bbm}
\usepackage{dsfont}
\usepackage{braket}
\usepackage{yfonts}
\usepackage[mathscr]{euscript}
\usepackage{siunitx}
\usepackage{booktabs}
\usepackage{changepage}
\usepackage{subcaption}
\usepackage{color}
\usepackage{url}
\usepackage{breqn}

\usepackage[utf8]{inputenc}

\usepackage{hyperref}

\usepackage{caption}

\usepackage{dcolumn}

\begin{document}

\begin{frontmatter}

\title{Numerical investigation of the generalized Jang equation coupled to conformal flow of metrics}

\author[mymainaddress,currentaddress]{Hollis Williams\corref{cor1}}

\address[mymainaddress]{%
Theoretical Sciences Visiting Program, 
Okinawa Institute of Science and Technology Graduate University \\ 
Onna-son, Okinawa 904-0495, Japan
}

\address[currentaddress]{%
Department of Mathematics and Statistics, University of Exeter, Exeter EX4 4QF, UK
}

\cortext[cor1]{Corresponding author. Email: holliswilliams@hotmail.co.uk}

\begin{abstract}\rm
\begin{adjustwidth}{2cm}{2cm}{\itshape\textbf{Abstract:}} 
A recent result of Jaracz has established nonexistence of global solutions to the coupled generalized Jang equation and zero divergence system which satisfy the asymptotic conditions needed to prove the Penrose conjecture by identifying a breakdown mechanism for the Jang slope at finite radius. In this work, we investigate whether a similar obstruction arises when the generalized Jang equation is instead coupled to the conformal flow of metrics.  Restricting to spherical symmetry and time-symmetric initial data, we formulate a numerically tractable version of the Jang/conformal flow system.  Our numerical results show no evidence of a finite radius breakdown analogous to that observed by Jaracz. Instead, the Jang slope remains regular and approaches its limiting value asymptotically. This behavior persists under controlled perturbations of the warping factor, indicating robustness of the observed phenomenon.  These findings suggest that coupling to conformal flow of metrics alters the obstruction mechanism present in the Jang/zero divergence system, and hence that this system may still be viable for proving the Penrose conjecture.
\end{adjustwidth}
\end{abstract}
\end{frontmatter}


\section{Introduction}

\subsection{The Penrose conjecture}

\noindent
The Penrose inequality relates the total mass of an asymptotically flat initial data set for the Einstein equations to the area of its outermost apparent horizon \cite{penrose}.  A popular form of the inequality states that
\begin{equation}
M_{\mathrm{ADM}} \geq \sqrt{\frac{A}{16\pi}},
\end{equation}
where $M_{\mathrm{ADM}}$ is the ADM mass and $A$ is the area of the outermost minimal surface, with saturation occurring for the Schwarzschild metric.  The Penrose conjecture states that equation (1) is true for any asymptotically flat initial data set $(M,g,k)$ satisfying suitable asymptotic boundary conditions and the dominant energy condition (in other words, for any physically reasonable initial data set).  A slightly different formulation of the conjecture is known to be true for spherically symmetric data sets (in this version, one considers the outermost future or past apparent horizon) \cite{hayward, bray1}.  The conjecture is also known in the important special case of time-symmetric data $(M,g)$, where $M$ is a Riemannian $3$-manifold.  The proofs of this statement both use geometric flows (either inverse mean curvature flow or conformal flow of metrics) \cite{huisken, bray2}.  The conjecture remains open in full generality.  One of the main challenges is the presence of extrinsic curvature, which prevents a direct reduction to the Riemannian case and complicates the application of methods from geometric analysis.  There have also been proofs of extensions of the Penrose inequality to include charge or angular momentum, but these invoke strong geometric assumptions similar to those of the Riemannian Penrose inequality \cite{disconzi, yamada, weinstein, jaracz2}.

\subsection{The generalized Jang equation}

 \noindent
The Jang equation was introduced in the 1970s as a potential method for proving the positive energy theorem \cite{jang}.  A full proof of this theorem with the Jang equation was provided by Schoen and Yau shortly afterwards \cite{schoen}.  The overall idea of the Jang equation is to produce a geometric perturbation which is designed to transform a general initial data set into one with improved scalar curvature properties. Given an initial data set $(M,g,k)$, the Jang perturbation seeks a graph hypersurface whose induced metric captures key features of the original data whilst allowing control of scalar curvature.  One can then complete the proof by reducing back to a proof of the positive energy theorem in the case of positive scalar curvature \cite{schoen2}. Malec and Ó Murchadha subsequently showed that despite this success, the Jang equation cannot be used to prove the Penrose inequality even in the spherically symmetric case where it reduces to a first order differential equation \cite{murchadha}.  The main technical problem was found to be lack of control of the ADM mass under conformal transformations.

In order to attempt a proof of the Penrose inequality in this approach, it is therefore necessary to generalize the Jang equation and avoid the obstructions that occur when trying to use the original version \cite{bray3}.  In the generalized version of the Jang equation, one searches for a hypersurface $\Sigma$ given by a graph $t=f(x)$ in the product space $M \times \mathbb{R}$, but now the product metric is a warped product metric $\overline{g} =  - \phi^2 d f^2 + g$ , where $\phi$ is a nonnegative function defined on $M$.  The Jang surface $\Sigma$ should satisfy a PDE of the form

\begin{equation} H_{\Sigma} = \text{Tr}_{\Sigma} K , \end{equation}

\noindent
where $H_{\Sigma}$ is the mean curvature of the hypersurface and $\text{Tr}_{\Sigma}K$ is the trace of the extrinsic curvature extended to the product $M \times \mathbb{R}$ and taken over $\Sigma$.  Define a symmetric 2-tensor $K$ on $M \times \mathbb{R}$ by extending $k$ as follows:

\[ K(\partial_{x^i}, \partial_{x^j}) = k_{ij} \: \: \text{for} \:\: 1 \leq i,j \leq 3, \tag{3a}\]

\[ K(\partial_{x^i}, \partial_t) = K(\partial_t, \partial_{x^i})  = 0 \: \: for \:\: 1 \leq i \leq 3, \tag{3b} \]

\[ K( \partial_t, \partial_t) = \frac{\phi^2 g(\nabla f, \nabla \phi)}{\sqrt{1 + \phi^2 |\nabla f|^2}}    , \tag{3c}\]

\noindent
where $x^i$ are local coordinates on $M$.  When $k$ is extended in this way, the generalized Jang equation in local coordinates is

\[ \Bigg( g^{ij} - \frac{\phi^2 f^i f^j}{1 + \phi^2 |\nabla f|^2} \Bigg) \Bigg( \frac{\phi \nabla_{ij}f + \phi_i f_j + \phi_j f_i}{\sqrt{1 + \phi^2 |\nabla f|^2}} - k_{ij} \Bigg) =0. \tag{4} \]

This modification alters the structure of the equation and is designed precisely to overcome the limitations of the original formulation.  In particular, coupling the equation to additional geometric data leads to solutions whose qualitative behaviour is fundamentally different.  There are several auxiliary equations which can be coupled to equation (4) to produce systems which can (in theory) be used to prove the Penrose inequality.  One of the simplest is the Jang-zero divergence system, where the auxiliary equation is simply $\overline{\text{div}}(\phi q) =0$.  This system admits solutions when linearized, but it was recently shown by Jaracz that the full system is not solvable even when spherical symmetry is assumed \cite{williams2, jaracz}.  It is also possible to produce viable systems by coupling to inverse mean curvature flow or conformal flow of metrics, but the analysis becomes considerably more complicated.  

\subsection{Jang/conformal flow system}

\noindent
In this work, we will consider the latter system.  The main idea of this system is to couple the generalized Jang equation to a conformal deformation of the metric.  To do this, one considers 
a one-parameter family of metrics
\[
g_t = u_t^4 g,  \tag{5}
\]
where $u_t > 0$ evolves by Bray's conformal flow of metrics.  The conformal factor $u_t$ satisfies
\[\partial_t u_t = v_t u_t , \tag{6}  \]
where $v_t$ is harmonic with respect to $g_t$:
\[\Delta_{g_t} v_t = 0 \quad \text{on } M_t . \tag{7} \]

\noindent
The boundary and asymptotic conditions for $v_t$ are

\[v_t = 0 \quad \text{on } \partial M_t , \tag{8a} \]
\[v_t(x) \to -1 \quad \text{as } |x| \to \infty .\tag{8b}\]

\noindent
In the zero scalar curvature formulation of Han and Khuri, the warping factor $\phi$ 
is determined via a scalar curvature correction \cite{han}.  We introduce a function $z_t$
satisfying
\[\Delta_{\bar g_t} z_t - \frac{1}{8} R_{\bar g_t} z_t = 0 ,  \tag{9}  \]
with asymptotic boundary condition
\[z_t \to 1 \quad \text{as } |x| \to \infty . \tag{10} \]

In this formulation, the warping factor $\phi$ 
admits an explicit integral representation along the conformal flow. Specifically,
$\phi$ is given by
\[
\phi(x)
=
2 \int_0^\infty \chi_t(x)
\left(
w_t^-(x)\, z_t^-(x)
+
w_t^+(x)\, z_t^+(x)
\right)
u_t^2(x)\, dt ,\tag{11} \]
where $\chi_t$ is an indicator function on $\Sigma_t$, 
$u_t$ is the conformal factor evolving under the conformal flow, $w_t^\pm$ denote the null expansions of the Jang surface, and $z_t^\pm$ are the scalar curvature correction factors arising from equation (9) on suitable ''mirror slices'' of $M_t$ defined in \cite{han}.  When an apparent horizon is present, the Jang function $f$ is required to blow up 
appropriately at $\partial M$, whereas the induced metric $\bar g_t$ remains regular.  At spatial infinity, the asymptotic boundary conditions are

\[f(x) \to 0, \quad \phi(x) \to 1 \quad \text{as } |x| \to \infty .  \tag{12}  \]

\noindent
Solutions for have been shown to exist for this system given restrictive assumptions, but a full existence theory is beyond current techniques \cite{williams1}.  There is also an alternative formulation for the system using spinors, but this runs into further subtleties involving spin geometry \cite{witten, taubes}.  A key issue with solvability is that the generalized Jang equation is degenerate and may fail to admit global solutions, with breakdown often occurring when the slope of the graph approaches a critical value.  Jaracz demonstrated that this degeneracy leads to nonexistence of solutions for the Jang/zero divergence system, even in spherical symmetry \cite{jaracz}. By identifying a finite radius blow-up mechanism for the Jang slope, that work effectively ruled out this particular formulation as a viable route toward the Penrose inequality.  Motivated by this result, it is natural to ask whether a similar nonexistence mechanism arises for alternative couplings to the generalized Jang equation.  

The goal of the present paper is to investigate numerically whether a Jaracz-type breakdown mechanism is likely to persist for the Jang/conformal flow system.  Restricting to spherical symmetry and time-symmetric initial data, we formulate a reduced system of ordinary differential equations capturing the essential coupling structure of the system. This setting allows for a detailed numerical study while retaining the key degeneracies responsible for potential nonexistence.  After validating our numerical framework against exact Schwarzschild data and verifying correct asymptotic and near-horizon behavior, we analyze the evolution of the Jang slope and its derivative. In contrast to the zero divergence system, we find no evidence of finite-radius breakdown. Instead, the slope approaches its limiting value asymptotically, and this behavior is robust under perturbations of the warping factor.  These results suggest that the conformal flow coupling alters the qualitative behavior of the Jang equation in a way which avoids the obstruction identified by Jaracz, providing numerical evidence that nonexistence arguments based on Jang slope blow-up do not directly extend to this system.

\section{Numerical formulation}

\noindent
We formulate a numerical implementation of the generalized Jang/conformal flow system. Throughout, we restrict attention to time-symmetric initial data so that the extrinsic curvature vanishes.

\subsection{Geometric setting}

\noindent
We assume that the the spatial metric is spherically symmetric and conformally flat, taking the form
\[g = u(r)^4 \left( dr^2 + r^2 d\Omega^2 \right),   \tag{13} \]

\noindent
where $u(r)$ is a positive conformal factor and $d\Omega^2$ denotes the standard metric on the unit sphere. In this setting, all geometric quantities reduce to functions of the radial coordinate $r$.  The initial horizon is located at $r=r_h$, where we impose regularity and normalization conditions consistent with the Schwarzschild solution. As $r \to \infty$, asymptotic flatness is enforced through suitable boundary conditions.  The conformal flow of metrics is generated by a harmonic function $v(r)$ satisfying
\[\Delta_g v = 0,  \tag{14}\]

\noindent
with boundary conditions

\[v(r_h) = 0, \qquad v(r) \to -1 \quad \text{as } r \to \infty.  \tag{15} \]

\noindent
In spherical symmetry, this equation reduces to a second-order ODE, which we solve numerically using finite differences.  The conformal factor is then defined by

\[u(r) = e^{v(r)}, \tag{16} \]

\noindent
consistent with the normalization of the Schwarzschild solution.

In spherical symmetry and for time-symmetric data, the scalar curvature equation reduces to a linear elliptic equation for an auxiliary function $z(r)$,
\[\Delta_{\tilde g} z - \frac{1}{8} R_{\tilde g} z = 0, \tag{17} \]

\noindent
with asymptotic boundary condition $z \to 1$ at infinity. For the data considered here, the numerical solution satisfies $z \equiv 1$ to machine precision, consistent with the Schwarzschild benchmark.  In the original system, the warping factor $\phi$ appearing in the generalized Jang equation is determined by a nonlocal integral representation along the 
conformal flow. This construction is designed to ensure 
vanishing scalar curvature of the Jang metric while coupling the Jang deformation 
to the conformal flow in a fully general setting.  In the present work, which is restricted to spherical symmetry and time-symmetric 
initial data, we adopt a simplified but compatible choice of warping factor by setting
\[\phi(r) = u(r),  \tag{18}  \]

\noindent
where $u$ is the conformal factor of the evolving metric. This choice agrees exactly 
with the Schwarzschild solution and satisfies all required regularity and asymptotic 
conditions. Moreover, in spherical symmetry, the conformal factor fully encodes the metric dependence, so that $\phi=u$ represents the most 
natural local realization of the warping factor consistent with the conformal flow.  This choice allows us to isolate and numerically 
investigate the behavior of the generalized Jang equation under a geometrically 
natural warping, without introducing nonlocal dependencies which would make the analysis intractable (even numerically). To assess the robustness of our conclusions with respect to 
this choice, we further consider controlled perturbations of $\phi=u$ and 
examine their effect on the Jang slope equation.

Finally, under the above assumptions, the generalized Jang equation for the Jang graph $f(r)$ reduces from a quasilinear elliptic PDE to a first-order ODE. Defining the normalized gradient
\[
Q(r) = \frac{f'(r)}{\sqrt{1 + f'(r)^2}},  \tag{19} \]
the Jang equation can be rewritten as an ODE for $Q(r)$,
\[
Q'(r) = F(r,Q,u,\phi).  \tag{20}  \]

\noindent
The explicit form of the function $F$ depends on derivatives of $u$ and $\phi$ and encodes the coupling between the Jang graph and the conformal flow of metrics.  The ellipticity of the original PDE degenerates as $|Q| \to 1$, so monitoring $Q(r)$ and its derivative $Q'(r)$ provides a sensitive diagnostic of potential breakdown at finite radius. Numerically, $Q$ is integrated outward from the horizon with initial condition $Q(r_h)=0$ and $Q'(r)$ is computed directly from the ODE to track the approach to the degeneracy barrier.

Our assumptions reduce the Jang/conformal flow system to a set of ODEs for a small number of scalar functions, including the conformal factor, the warping factor, and the Jang slope.  Although this reduction greatly improves tractability, the resulting equations remain nonlinear and degenerate, and their qualitative behavior is not a priori clear.  Importantly, spherical symmetry does not trivialize the problem of existence. Previous work has shown that even in this setting, the Jang/zero divergence system exhibits finite-radius breakdown \cite{jaracz}.  Our results are therefore not intended as a proof of the Penrose inequality, but rather as numerical evidence that a previously noted nonexistence mechanism likely fails for the coupled Jang/conformal flow system, even in a highly symmetric setting.

\subsection{Numerical implementation}

\noindent
All equations are discretized on a one-dimensional radial grid $r \in [r_h, R_{\max}]$ with uniform spacing. The outer boundary $R_{\max}$ is chosen to be sufficiently large that asymptotic boundary conditions are accurately approximated.  Convergence with respect to $R_{\max}$ was verified as part of the numerical validation.  Second-order finite difference schemes are used to approximate radial derivatives. Elliptic equations arising from the conformal flow and scalar curvature constraint are solved using standard finite difference discretizations, leading to sparse linear systems which are solved using direct sparse solvers. Boundary conditions at the horizon and at the outer boundary are imposed directly at the grid endpoints.  The ODE for the Jang slope $Q(r)$ is integrated outward from the horizon using an explicit first-order scheme with step size equal to the grid spacing.  Regularity at the horizon is enforced by prescribing the initial condition $Q(r_h)=0$.  Its derivative $Q'(r)$ is computed directly from the slope equation.  To assess numerical stability and robustness, computations were repeated at multiple grid resolutions, and all qualitative features of the solutions were found to be resolution-independent. In particular, the behavior of the Jang slope near its limiting values was stable under mesh refinement.  Controlled perturbations of the warping factor $\phi$ were also tested, confirming that the system's behavior is insensitive to small deviations from the exact choice.

\section{Validation}

\noindent
To validate the numerical implementation, we benchmark the harmonic function solver against the exact Schwarzschild solution. Table 1 presents a mesh refinement study at fixed outer boundary radius $R_{\text{max}}$.  The errors remain approximately constant under refinement, indicating that the dominant contribution arises from finite-domain truncation associated with imposing asymptotic boundary conditions at a finite radius.  To isolate truncation effects, we perform a second study in which the outer boundary radius $R_{\text{max}}$ is increased while maintaining proportional grid resolution. As shown in Table 2, both the $L^{\infty}$ and $L^2$ errors decrease as $R_{\text{max}}$ is increased, demonstrating convergence of the numerical solution to the exact Schwarzschild harmonic function in the asymptotically flat limit.  These benchmarks establish the accuracy and robustness of the numerical framework used in the subsequent investigation of the coupled generalized Jang and conformal flow system.  Fig. 1 shows the numerical solution of $\Delta v = 0$ plotted against the analytic solution for Schwarzschild initial data.  Throughout the benchmark, we work in geometric units with mass parameter $M =1$, so that the horizon is located at $r_h = 2$.

\begin{table}[h]
\centering
\begin{tabular}{lcc}
\hline
$N$& $L^{\infty}$ error & $L^2$ error \\
\hline
200 & $1 \times 10^{-2}$& $1.362870 \times 10^{-1}$ \\
400 &  $1 \times 10^{-2}$ & $1.361961 \times 10^{-1}$\\
800 &  $1 \times 10^{-2}$& $1.361505 \times 10^{-1}$ \\
1600 &  $1 \times 10^{-2}$& $1.361277 \times 10^{-1}$ \\
\hline
\end{tabular}
\caption{Mesh refinement study for the Schwarzschild harmonic function benchmark at fixed outer boundary radius $R_{\text{max}}$. The $L^{\infty}$ and $L^2$ errors remain approximately constant under refinement, indicating that the dominant contribution arises from finite-domain truncation rather than discretization error.}
\label{tab:results_comparison}
\end{table}

\begin{table}[h]
\centering
\begin{tabular}{lccc}
\hline
$R_{\text{max}}$& $N$  &  $L^{\infty}$ error & $L^2$ error \\
\hline
100 &  400 &  $2 \times 10^{-2}$& $1.875219 \times 10^{-1}$ \\
200 & 800  &$2 \times 10^{-2}$& $1.361505 \times 10^{-1}$  \\
400 & 1600  &   $5 \times 10^{-3}$ &$9.781842 \times 10^{-2}$  \\
\hline
\end{tabular}
\caption{Convergence study with increasing outer boundary radius $R_{\text{max}}$ and proportional grid refinement. Both $L^{\infty}$ and $L^2$  errors decrease as $R_{\text{max}}$ is increased, demonstrating convergence of the numerical solution to the analytic Schwarzschild harmonic function in the asymptotically flat limit.}
\label{tab:results_comparison}
\end{table}

We also compute derived quantities from $v(r)$, including the conformal factor $u(r) = e^{v(r)}$, the zero scalar curvature correction $z(r) =1$, the warping factor $\phi(r) = u(r)$, and the Jang slope $Q(r) = 0$. Diagnostics (max/min/absolute errors) confirm that these quantities are consistent with their analytic values across the computational domain. Since all of these are directly derived from the harmonic function, separate convergence studies for these quantities were not performed.  Near-horizon diagnostics were performed by examining the numerical solution in a neighborhood of $r = r_h$.  While the specific grid points differ under refinement due to uniform discretization of the domain, the qualitative and quantitative behavior of the solution near the horizon is stable (shown in Fig. 2). In particular, the horizon values  $u(r_h) =1$, $\phi (r_h) = 1$, and $Q(r_h) = 0$ are preserved under grid refinement, and no spurious boundary layers or oscillations are observed.  For Schwarzschild data, the warping factor satisfies $\phi(r) = u(r) $ and the Jang slope satisfies $Q(r) =0$ to numerical precision.  These properties were verified directly and are therefore not plotted separately.  Mesh refinement studies demonstrate convergence consistent with the expected second-order accuracy of the finite difference discretization.

\begin{figure}
\centering
\includegraphics[width=100mm]{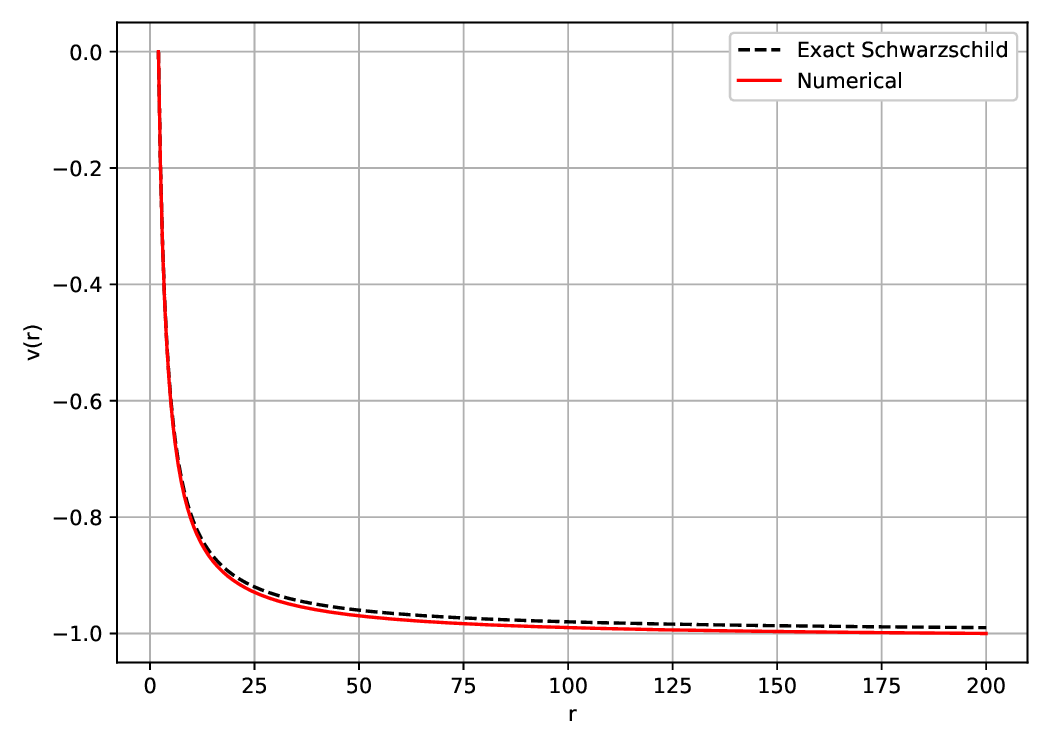}%
\caption{\label{fig:epsart} Numerical solution of the harmonic equation $\Delta v = 0$ in spherical symmetry compared with the exact Schwarzschild solution $v(r) = -1 + r_h/r$.  The numerical solution is obtained using a second-order finite difference discretization with horizon and asymptotic boundary conditions. Excellent agreement is observed across the domain, validating the numerical implementation of the conformal flow equation. }
\end{figure}

\begin{figure}
\centering
\includegraphics[width=100mm]{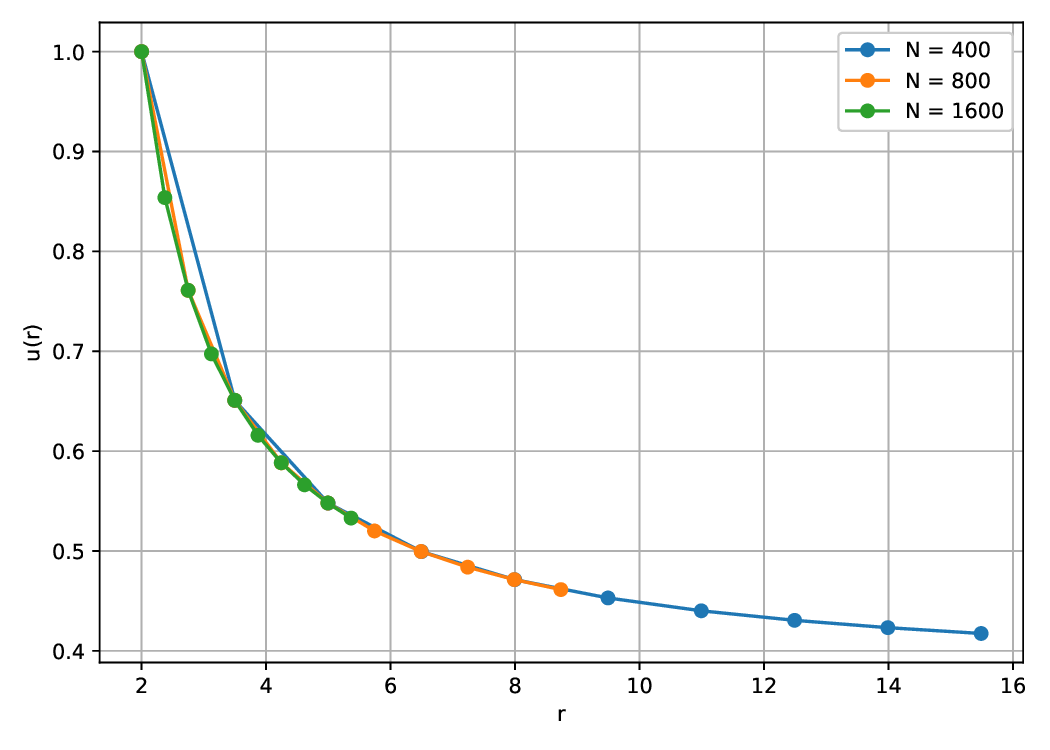}%
\caption{\label{fig:epsart} Near-horizon behavior of the conformal factor $u(r)$ for Schwarzschild initial data at increasing resolution $N$. The numerical solutions satisfy the expected horizon condition $u(r_h) =1$.  Although the curves do not coincide pointwise near the horizon due to differing grid spacings, the limiting behavior as $r \rightarrow r_h^+$ is stable under refinement and no spurious boundary layers are observed.}
\end{figure}

\section{Results}

\subsection{Avoidance of the Jaracz breakdown mechanism}
\label{subsec:jaracz_avoidance}

\noindent
In spherical symmetry, we define a variable $Q(r)$ which we call the Jang slope:
\[
Q(r) = \frac{f'(r)}{\sqrt{1 + (f'(r))^2}} \,,  \tag{21}
\]
which is equivalent to the derivative of the Jang graph in isotropic coordinates, up to a monotone reparameterization that maps the slope to the bounded interval $[-1,1]$.  Jaracz showed that for the Jang/zero divergence system, global solutions generically fail to exist due to a breakdown of the Jang slope at finite radius. In spherical symmetry this manifests as blow-up of the slope variable \(Q\) as \(|Q| \to 1\), where the reduced equation loses ellipticity and the solution cannot be extended further.  Although Jaracz did not explicitly define the same variable $Q$, his analysis of finite radius blow-up of $f'(r)$ in spherical symmetry is equivalent.  The variable $Q = f'/\sqrt{1 + (f')^2}$ captures the same ellipticity breakdown, with $|Q| \to 1$ corresponding to divergence of the Jang graph.  In \cite{jaracz}, the blow-up of the radial derivative $f'(r)$ was analyzed directly. Tracking $Q$ numerically allows a convenient measure of ellipticity loss: as $|Q| \to 1$, the Jang graph diverges.  This reformulation allows us to track the blow-up in a numerically stable manner while preserving the analytic significance of the breakdown.

We investigate this mechanism numerically for the spherically symmetric Jang/conformal flow system. In this formulation, the Jang equation is coupled to a dynamically determined warping factor \(\phi\), which is itself linked to the conformal factor through the zero scalar curvature condition.  This yields an evolution equation for the Jang slope of the form
\[
Q'(r)
=
\left(1 - Q(r)^2\right)
\left(
\frac{2}{r}
+
\frac{\phi'(r)}{\phi(r)}
\right),
\label{eq:Q_ode} \tag{22} \]
with initial condition \(Q(r_h)=0\) imposed at the horizon.  Under the assumptions of spherical symmetry, time symmetry, and conformal flatness, all angular derivatives vanish and the generalized Jang equation reduces to a first-order ODE for the normalized slope $Q$, with all geometric dependence encoded into the conformal factor and warping function.  We solve equation \eqref{eq:Q_ode} numerically using the same discretization and solver settings validated against Schwarzschild initial data in Section 3, with results plotted in Figs. 3 and 4. For all parameter choices tested, the numerical solutions exhibit the following behavior:
\begin{itemize}
\item \(Q(r)\) increases monotonically from zero and approaches \(|Q|=1\) only asymptotically.
\item The derivative \(Q'(r)\) remains bounded for all finite \(r\) and decays to zero as \(|Q|\to 1\).
\item No finite--radius singularity or loss of regularity is observed up to the largest computational radius.
\end{itemize}

\noindent
This behavior is qualitatively different from the Jaracz breakdown mechanism. In the Jang/zero divergence system, the slope equation becomes singular as \(|Q|\to 1\), forcing blow-up at finite radius. In contrast, the conformal flow coupling introduces a damping factor \((1-Q^2)\) multiplying the right-hand side of equation \eqref{eq:Q_ode}, converting \(|Q|=1\) from a singular barrier into a degenerate but stable asymptotic state.  Numerically, this is reflected in the saturation of \(Q(r)\) at \(|Q|=1\) together with the decay of \(Q'(r)\) to zero, rather than divergence. This suggests that the conformal flow formulation avoids the obstruction identified by Jaracz, at least within the spherically symmetric setting studied here.

\begin{figure}
\centering
\includegraphics[width=100mm]{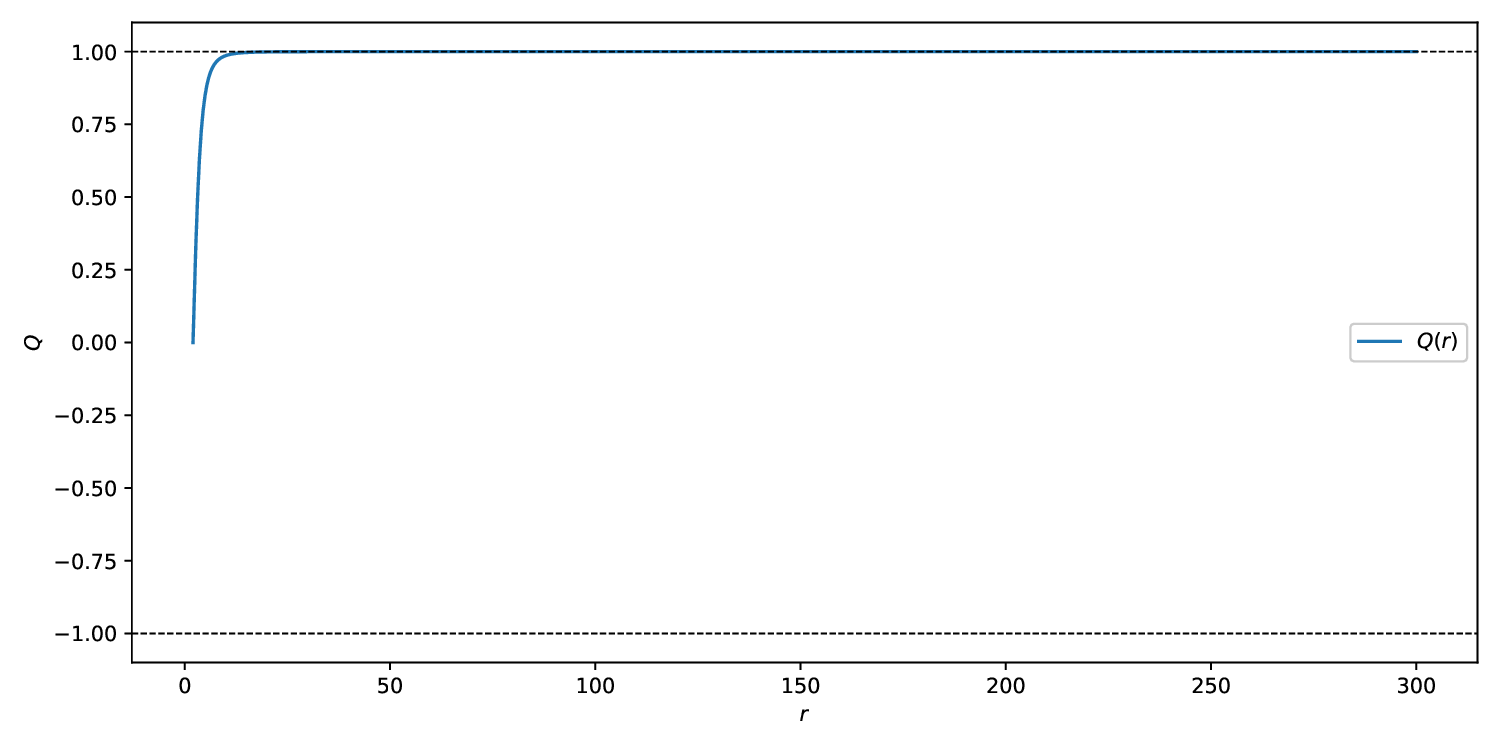}%
\caption{Numerical solution for the Jang slope \(Q(r)\) in the spherically symmetric conformal flow system. The slope increases monotonically from the horizon value \(Q(r_h)=0\) and approaches the limiting value \(|Q|=1\) only asymptotically. No finite radius breakdown is observed, in contrast to the Jang/zero divergence system.
}

\end{figure}

\begin{figure}
\centering
\includegraphics[width=100mm]{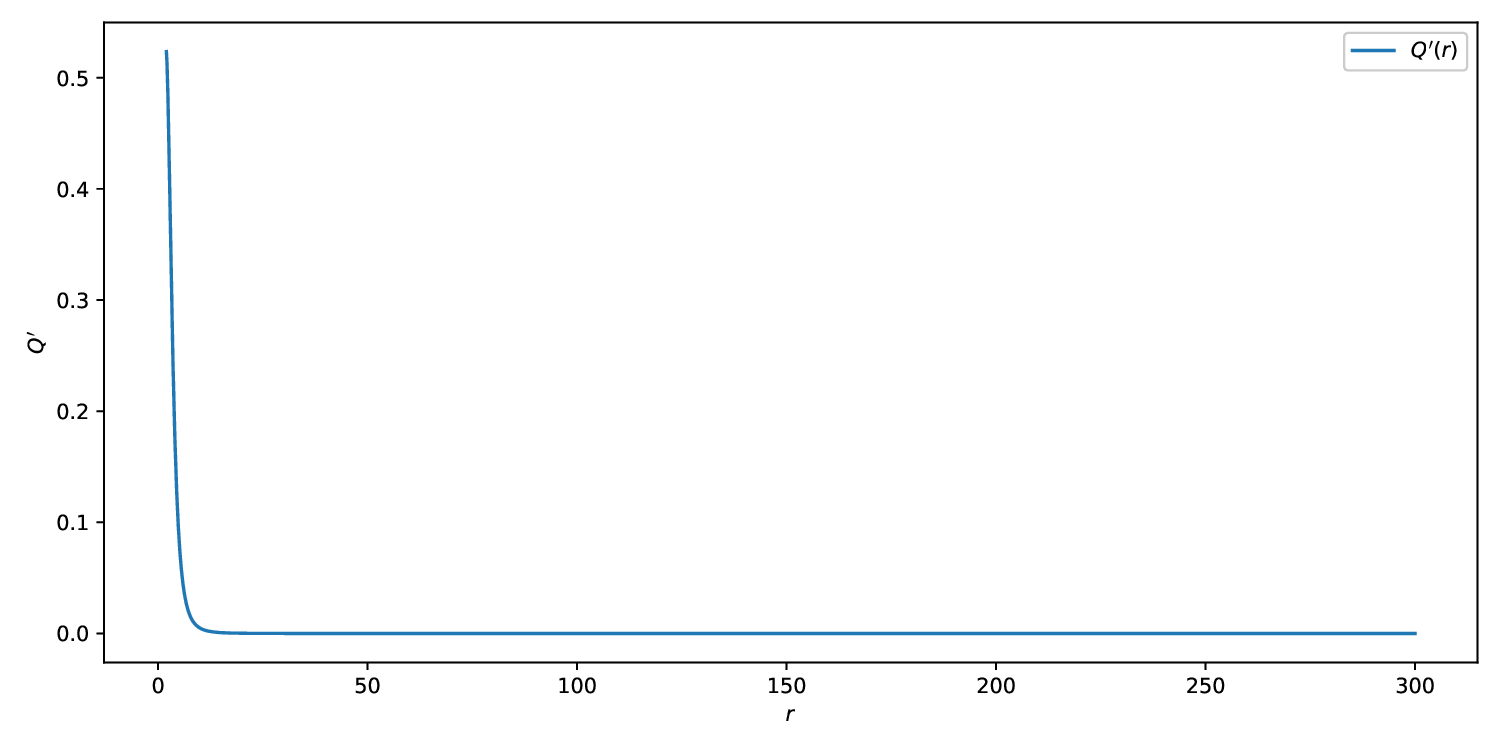}%
\caption{Radial derivative \(Q'(r)\) corresponding to the solution shown in Fig. 3. The derivative remains bounded and decays to zero as \(|Q|\to 1\), indicating saturation of the Jang slope, rather than blow-up at finite radius.
}

\end{figure}

\subsection{Robustness under perturbations of the warping factor}
\label{subsec:phi_robustness}

\noindent
We next investigate whether the evasion of the Jaracz breakdown mechanism observed in Section 4.1 is sensitive to the specific choice made for $\phi$. To test this, we introduce controlled perturbations of the warping factor whilst keeping the background geometry fixed.  Starting from the baseline warping factor $\phi_0(r)$ obtained from the conformal flow, we consider the one-parameter family
\[
\phi_\varepsilon(r)
=
\phi_0(r)\,\bigl(1 + \varepsilon\,p(r)\bigr),   \tag{23}\]
where $\varepsilon \in \mathbb{R}$ and $p(r)$ is a smooth positive function. In our numerical experiments we take
\[
p(r) = e^{-(r-r_h)/L}, \tag{24} \]
with decay length $L>0$, so that the perturbation is localized near the horizon and decays exponentially at infinity. This choice preserves the regularity of $\phi$ at the horizon as well as asymptotic flatness.

For each value of $\varepsilon$, we solve equation (5) using the same numerical discretization validated in Section 3.  Across a wide range of perturbation strengths, including $|\varepsilon|\leq 0.5$, we observe no finite radius breakdown of the Jang slope. In all cases, $Q(r)$ increases monotonically from the horizon value and approaches the limiting value $|Q|=1$ only asymptotically, whilst the derivative $Q'(r)$ remains bounded and decays to zero at large radius (shown in Figs. 5 and 6. Quantitatively, the maximum value of $|Q'(r)|$ varies smoothly and monotonically with $\varepsilon$, but no qualitative change in behavior is observed.  This provides evidence that the avoidance of the Jaracz breakdown mechanism is not an artifact of a finely tuned choice of $\phi$, but rather a robust feature of the conformal flow coupling.

\begin{figure}
\centering
\includegraphics[width=100mm]{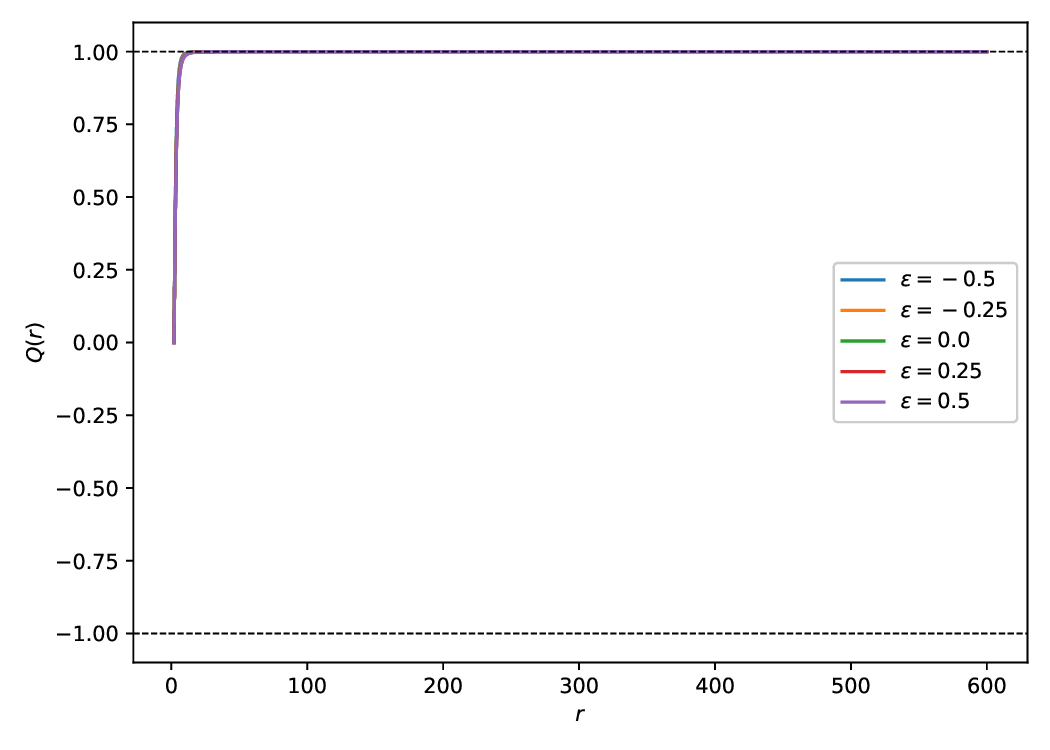}%
\caption{Jang slope $Q(r)$ for the perturbed warping factors $\phi_\varepsilon$ defined in equation (6), shown for several values of $\varepsilon$. All solutions exhibit saturation at $|Q|=1$ only asymptotically, with no finite radius breakdown.
}

\end{figure}

\begin{figure}
\centering
\includegraphics[width=100mm]{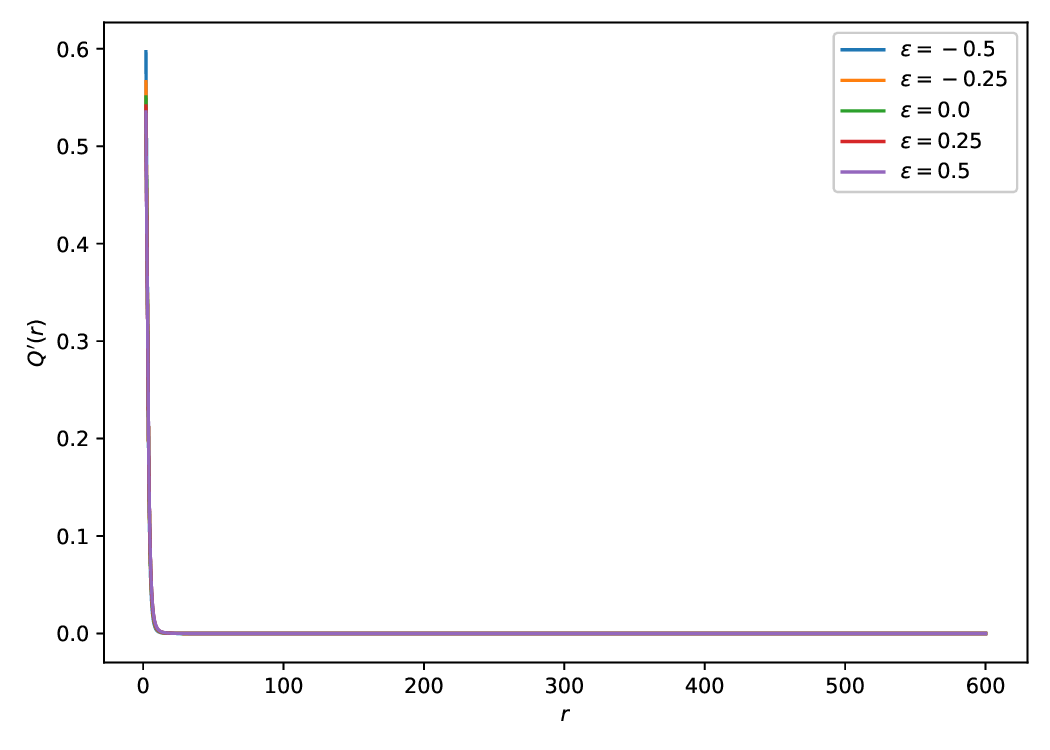}%
\caption{Radial derivative $Q'(r)$ corresponding to the solutions in Fig. 5. The derivative remains bounded for all perturbation strengths and decays to zero as $|Q|\to 1$, indicating saturation rather than singular behavior.
}

\end{figure}

\subsection{Perturbations of initial data}

\noindent
In the numerical experiments presented here, we have restricted to perturbations of the warping factor $\phi$ whilst keeping the underlying initial data fixed. In particular, the extrinsic curvature $k$ is not perturbed.  Perturbing $k$ generically alters the constraint equations and would requires a reformulation of the coupled Jang/conformal flow system, including modifications to the scalar curvature equation and the associated asymptotic boundary conditions. A systematic study of such perturbations would therefore require a substantially different numerical framework which is beyond the scope of this work.  We restrict here to testing the robustness of the Jang slope equation and its associated barrier structure under controlled variations of the warping factor within the same class of initial data. The results indicate that the qualitative behavior of the solutions, including the absence of finite radius breakdown and the asymptotic approach to the degeneracy barrier $|Q|=1$, is insensitive to such perturbations.

\section{Discussion}

\noindent
The numerical results presented above indicate a qualitative difference between the behavior of the Jang/conformal flow system and the breakdown mechanism identified by Jaracz for the coupled Jang/zero divergence system. In this section, we discuss the origin of this difference, clarify its significance, and summarize the implications of our findings.

\subsection{Failure of the Jaracz nonexistence mechanism}

\noindent
Jaracz’s nonexistence argument relies on the emergence of a finite radius degeneracy in the Jang slope variable $Q$, driven by the structure of the zero divergence condition and its decoupling from any geometric evolution which could compensate for the divergence. In that setting, the slope equation admits trajectories which reach the barrier $|Q|=1$ at finite radius, leading to loss of regularity and breakdown of the solution.  In contrast, in the present system the Jang equation is coupled to a conformal deformation of the metric through the warping factor $\phi$. Even in spherical symmetry, this coupling alters the effective coefficients in the slope equation. Numerically, we observe that although $Q(r)$ approaches the degeneracy barrier $|Q|=1$, it does so only asymptotically, with no indication of finite radius blow-up or loss of regularity.  At the same time, the derivative $Q'(r)$ decays to zero, indicating stabilization of the slope dynamics rather than instability.  These observations provide numerical evidence that the conformal flow coupling suppresses the finite radius breakdown mechanism present for the Jang/zero divergence system. From the perspective of the reduced slope equation, the warping factor modifies the balance between geometric forcing and barrier formation so that trajectories are attracted to, but do not cross, the degeneracy threshold.

\subsection{Role of time symmetry}

\noindent
All numerical experiments presented here are performed under the assumption of time-symmetric initial data, so that the extrinsic curvature $k$ vanishes identically. Although this simplifies the structure of the equations, we emphasise that it does not trivialize the problem. Even in this setting, the generalized Jang equation remains nonlinear and degenerate, and its coupling to the conformal flow of metrics plays a decisive role in the observed behavior.  Importantly, the absence of finite radius breakdown is not a purely kinematic consequence of setting $k=0$. In the Jang/zero divergence system studied by Jaracz, breakdown occurs even in spherical symmetry. The present results therefore suggest that it is the conformal flow coupling, rather than time symmetry alone, which alters the qualitative behavior of the slope equation.  Perturbations of the extrinsic curvature would require a reformulation of the coupled system, including modifications to the scalar curvature equation and its asymptotics, and are therefore left for future work.  The robustness of the observed behavior under controlled perturbations of the warping factor nevertheless provides evidence that the absence of finite radius breakdown is intrinsic to the coupled Jang/conformal flow system, rather than an artifact of exact symmetry or fine-tuning.

\subsection{Implications and outlook}

\noindent
The numerical experiments presented here suggest that proofs of nonexistence which rely on forcing the Jang slope to reach the degeneracy barrier at finite radius are unlikely to apply directly to the Jang/conformal flow system, at least within the class of data considered here. Instead, the system exhibits asymptotic saturation of the slope variable, with solutions extending smoothly to large radius.  These results do not constitute a proof of global existence. However, they provide numerical evidence that a natural and previously successful nonexistence mechanism fails for the coupled Jang/conformal flow system, and so provide evidence that this system may still be a correct approach for proving the conjecture once the necessary theory is developed.  This is similar to other works in the literature which provide numerical evidence for feasibility of other formulations of the Penrose inequality without constituting a proof \cite{karkowski, kulc}.  Taken together, our findings support the viability of the Jang/conformal flow approach as a framework which merits further analytic and numerical investigation in connection with the Penrose inequality.

\section*{Acknowledgments}

\noindent
HW thanks Jaroslaw Jaracz for useful discussions.  This research was partly conducted whilst HW was visiting the Okinawa Institute of Science and 
Technology (OIST) through the Theoretical Sciences Visiting Program (TSVP).  HW acknowledges support from a London Mathematical Society Early Career Research Travel Grant (ECR-2526-51).   

\section*{Data availability}

\noindent
The numerical codes used to generate the results presented in this article are
available at \url{https://github.com/Tom467/JangCFM/}. No experimental data
were generated or analysed during the current study.

 \section*{Conflict of Interest}

\noindent
The authors have no conflicts to disclose.


\end{document}